\documentclass[11pt,twoside]{article}

\usepackage{asp2004}
\usepackage{lscape}

\newcommand{\kms}{\ensuremath{{\rm km\,sec^{-1}}}}                   
\newcommand{\msunyr}{\ensuremath{\mathit{M}_{\odot}{\rm yr}^{-1}}}   

\newcommand{\mdot}{\ensuremath{\dot{M}}}                             

\begin{document}
\title{Mass loss summary -- a personal perspective}    
\author{Jorick S. Vink}   
\affil{Keele University, Astrophysics Group, Lennard-Jones Labs, ST5 5BG, Staffordshire, UK}    

\begin{abstract} 
For the occasion of the official retirement of Henny Lamers, a meeting was held to celebrate Henny's 
contribution to mass loss from stars and stellar clusters. 
Stellar mass loss is crucial for understanding the life and death of massive stars, 
as well as their environments. Henny has made important contributions to many aspects
of our understanding of hot-star winds. Here,
the most dominant aspects of the stellar part of the meeting: (i) O star wind clumping,  
(ii) mass loss near the Eddington limit, and (iii) and the driving of Wolf-Rayet winds, are highlighted.  
\end{abstract}


\section{Introduction}

Andre Maeder opened the meeting with a comprehensive 
review describing the basic framework for massive star evolution models, 
emphasizing the importance of mass loss and rotation. 
Henny has played a crucial role in 
our appreciation for the role of mass loss in stellar evolution. 
Within the canonical framework, mass loss ``peels off'' the outer envelope, and 
O stars evolve into Luminous Blue Variable (LBV), and Wolf-Rayet (WR) stars, before 
they explode as supernovae (SNe). 

Massive star winds became a hot research topic at the very 
start of Henny's illustrious career. The field of stellar winds
took off with the discovery of P Cygni profiles in the spectra 
of Orion's O stars \citep{morton67}, which initiated the construction of the 
first radiation-driven wind models by \cite{lucy70} and 
Castor, Abbott \& Klein (1975; CAK, thoroughly 
reviewed by John Bjorkman during this meeting), but it also 
motivated \cite{lamers76} to
determine the first accurate empirical mass-loss rates. 
Intriguingly, the mass-loss rate for the prototypical O star 
$\zeta$ Pup (not Pub!) of \mdot $=$ 7.2 $\times$ $10^{-6}$ \msunyr\ is  
very close to the current estimate.

This brings me to the main topic of this meeting: 
what are the {\it real} mass-loss rates of massive stars? 
After a discussion on the topic of wind clumping 
(Sect.~2), I will discuss the issues of mass loss in close proximity 
to the Eddington/Omega limit (Sect.~ 3), and the driving of Wolf-Rayet 
winds (in Sect.~4). I will focus on just a number   
of aspects of massive star mass loss and I hope I will be forgiven
for not attempting the impossible task of highlighting all topics in 
the limited 
amount of time and space available. 

\section{What are the {\it real} mass-loss rates?}

I borrow this title from Joachim Puls' contribution, as it was one of the 
main topics of this meeting. The background is the following: 
massive star evolutionary models either employ empirical or theoretical 
mass-loss formulae.
But over the last couple of years, there has been a growing number of studies, 
based on both X-ray [in these proceedings see 
Oskinova; Albacete-Colombo] and UV [Massa] data which indicate that O star
winds are significantly clumped, and canonical empirical mass-loss rates 
overestimated. Of course, no-one doubts that O star winds are clumped, but
the question that needs to be addressed is {\it by what factor} empirical and 
theoretical rates -- based on smooth line-driven wind models, might be in error. 
Even more fundamental is the question whether the effects of overestimating the mass-loss rate 
could significantly affect our understanding of massive star evolution. 

As was comprehensively discussed by Alex de Koter, the UV mass-loss 
rates are the most sensitive diagnostic, but the drawback is that
the wind ionization is very model dependent. Measuring the free-free radio
or H$\alpha$ emission leads to empirical mass-loss rates which are far less 
model dependent and probably more accurate 
\citep[e.g.][]{lamers93}. These radio and H$\alpha$ diagnostics however 
are both $\rho^{2}$ diagnostics and could be significantly affected by 
wind clumping. Progress can be made by investigating the radial dependence of the 
clumping factor \citep[e.g.][]{runacres02}. 

Puls and co-workers studied the empirical clumping stratification in the radial direction and 
they found the inner part of the winds (where the H$\alpha$ is formed) to be more strongly 
clumped than the outer wind regions (where the radio free-free emission is 
formed [Blomme]). If the outer wind is not significantly clumped, the evolutionary effects
may be modest.
If however, the outer winds are strongly clumped, the inner winds may be even more 
strongly clumped, and effects on our stellar models may be severe.
In short, until we know the clumping factor 
of the outer wind, we will not be able to provide a clear-cut answer to the 
crucial question ``what is the real mass-loss rate?''

To make progress, we need to put the pieces together. 
In addition to the X-ray, UV, H$\alpha$, infrared and radio diagnostics, 
probing the wind emission at different positions in the wind, polarimetry studies [Davies] 
may provide strong geometric constraints on wind structure.
Polarimetry results for LBVs call for an onset 
of wind clumping very close to the stellar photosphere, as to be able to produce the observed
levels of polarisation. Possibly, this situation may be the same for O stars.

In comparing observational aspects to our theoretical knowledge, 
I would like to take the opportunity to stress that a modest amount of wind clumping is
actually {\it required} to match the theoretical line-driven wind models. 
\cite{repolust04} for the case of Galactic supergiants, and \cite{mokiem07} 
for Magellanic Cloud supergiants needed to down-revise the empirical H$\alpha$ mass-loss rates by 
a factor 2-3, corresponding to a clump filling factor $f$ of $\sim$5, to match the 
theoretical wind-momentum luminosity relation. 

Whether clumping would significantly effect the theoretical models is as yet an open issue. 
On the one hand, one might expect photons to be trapped in clumps, 
thereby potentially limiting the multi-scattering process, but on the other hand, 
the photons might escape more easily into inter-clump medium, thereby travelling
larger distances and depositing their momentum more efficiently. 
An additional effect might be that clumping may lead to the recombination of the dominant line-driving 
ions, and a similar effect as that of ``bi-stability'' (see below) may actually lead to 
an {\it increase} rather than a decrease of the predicted mass-loss rate. 
Only the future will tell.  

\section{Mass loss in close proximity to the Eddington/Omega Limit}

Mass loss in close proximity to the Eddington limit for LBVs \citep[e.g.][]{lamers88} 
may be even more relevant for massive star evolution if the line-driven winds of O stars would turn
out to be insufficient in removing most of the stellar mass.  
Continuum-driven winds were extensively reviewed by Stan Owocki and 
their implications for Eta Car [Smith; Hillier; Gull] and other LBVs led 
to some heated discussions during the meeting. 

Furthermore, the effects of stellar rotation on the 
mass-loss rate [Owocki; Langer; Ceniga] were equally relevant. 
One of the most interesting discoveries was the broad 
Si {\sc iv} line of the LBV AG~Car [Groh], implying a projected rotational 
velocity of $\sim$ 200 \kms -- a significant fraction of the break-up speed.
This may confirm that LBVs reside close to the Omega-Eddington limit.

Stellar rotation may lead to wind asymmetries [Maeder]. 
For most objects, the oblate distortion of the star and Von Zeipel gravity
darkening for radiative envelopes may lead to a distribution of mass flux
which is stronger towards the pole than to the equator [Owocki]. 

This configuration of a polar mass flux distribution 
may be the exact opposite for objects that find themselves at temperatures 
in close proximity of the so-called ``bi-stability jump'' 
(Pauldrach \& Puls 1990, Lamers et al. 1995). 
At lower temperatures, the dominant line-driving ion of iron (Fe) 
recombines from {\sc iv} to {\sc iii}, which is predicted to  
result in larger mass flux from the equator than at the pole.
This ``rotationally enhanced bi-stability'' effect from \cite{lamers91} 
may be very relevant in explaining the B[e] phenomenon.

During this meeting, the first empirical evidence for the possible existence 
of a {\it mass-loss} bi-stability jump in the HRD was presented by 
Benaglia -- through a radio survey over the bi-stability range.
However, there remain a number of later type B supergiants for which
the empirical and theoretical rates are discrepant [Crowther]. 
Future work should show whether these B-star discrepancies may be 
related to wind clumping -- the main contender for the discordance 
of the UV mass-loss studies for O supergiants [Massa].

\section{The driving of Wolf-Rayet winds}

One of the most significant differences between Wolf-Rayet and O star winds is  
that WR stars invariably show very large ``pseudo-photospheres''. 
This makes it particularly challenging to study the innermost regions of 
WR winds. Despite this disadvantage, a lot of progress to our understanding
of the driving of WR winds has been made in the last couple of years. 

This progress happened on the purely analytic front [Onifer], the semi-analytic 
front, as well as the quantitative modelling front. The postulation of 
\cite{nugis02} that the continuum (OPAL) opacities play an instrumental role 
in starting the WR winds at deep layers helped the subsequent 
PoWR modellers [Gr\"afener, Hamann] to initiate and maintain the driving of a carbon-rich
WC star from the photosphere into the regime where the wind reaches 
its terminal velocity.  

This suggests that despite the differences with O star winds,
both groups of winds are primarily driven by radiation pressure. As a result, 
we are now able to estimate mass-loss dependencies of WR stars as a function of 
metal content [Vink, de Koter], with important consequences for gamma-ray burst progenitors and early 
generations of massive stars. 

Despite the successes, Gr\"afener warned that there are still many 
aspects of WR driving that remain poorly understood. One such aspect concerns the 
enigmatic WN8 stars, which are known to be highly variable. 
Observations with the MOST ``humble space telescope''
[Moffat] may tell us more about the role of WR pulsations.
Another potentially most relevant physical ingredient involves 
the strength of the surface magnetic field. During her talk, Nicole St.-Louis set an 
upper limit of 25 Gauss on at least one WR star.

Finally, I wish to highlight the role of rotation for WR stars. In the case of WR1, 
Chen\'e managed to derive a period of 16.7 days. Assuming a WR radius of 5 solar radii,  
the rotation speed at the stellar surface was estimated to be in the range 15-100 \kms [St. Louis].
Despite the necessary assumptions, this is clearly a unique constraint,
as WR rotation rates cannot be determined from the more traditional method of measuring 
$v$ sin$i$ from stellar absorption line 
spectra, due to the broad lines in their spectra.

\section{Final words}

One of the fiercest discussions during the meeting 
was related to the possibility that giant LBV eruptions ($\eta$ Car type eruptions, not
the typifying S Doradus variations) may be a dominant mechanism for the integrated 
mass loss during the life of a massive star, as severe clumping in O-star winds may 
imply negligible mass loss through line-driven
winds. 

Although the issue of the clumping factor in O star winds is very much an 
open one, if clumping factors would be significantly larger than $\sim$five, 
one might require extra mass loss during the LBV phase to 
produce the much lower masses of WR stars. An alternative could 
be an early SN explosion, as was discussed by several speakers [Smith; Vink].
Now that LBVs have been suggested as potential SNe progenitors, whilst e.g. Langer
suggested the possibility of quasi-homogeneous evolution of GRB progenitors, 
it appears that the successful ``standard scenario'' for 
massive star evolution may need some revision. 

Future mass-loss studies will play a major role in constructing 
a more complete picture of massive star evolution. These studies will no-doubt 
heavily rely on the insight of physical processes developed by Henny Lamers 
and his contemporaries. 

\acknowledgements 
Henny many thanks for sharing your deep knowledge and insight in physics and 
your unlimited enthusiasm for astronomy.

\end{document}